\documentclass[10pt, onecolumn]{IEEEtran}

\usepackage{algorithm}
\usepackage[noend]{algpseudocode} 
\usepackage{algorithmicx}

\usepackage{amsmath,amssymb,amsfonts,dsfont}
\usepackage{graphicx}

\usepackage{subfigure}
\usepackage{array,multirow,graphicx}

\usepackage{textcomp}
\usepackage{hyperref}
\usepackage{xcolor}
\newcommand*{\red}{\textcolor{red}}

\usepackage{lipsum}
\usepackage{epsfig}
\usepackage{float}
\usepackage{relsize}
\usepackage{comment}
\usepackage[shortlabels]{enumitem}
\usepackage{listings}
\usepackage[nocomma]{optidef}
\newcommand\scalemath[2]{\scalebox{#1}{\mbox{\ensuremath{\displaystyle #2}}}}
\usepackage{caption}
\usepackage{subcaption}
\usepackage{nopageno}
\usepackage{tabularx}
\usepackage{makecell}

\def\BibTeX{{\rm B\kern-.05em{\sc i\kern-.025em b}\kern-.08em
    T\kern-.1667em\lower.7ex\hbox{E}\kern-.125emX}}

\hypersetup{nolinks=true} 

\begin{document}

\title{Reinforcement Learning-driven Data-intensive Workflow Scheduling for Volunteer Edge-Cloud\thanks{This material is partially supported by the National Science Foundation under Award Numbers: OAC-2232889 and CNS-1943338.}}

\author{
Motahare Mounesan$^\ast$, Mauro Lemus$^\$$, Hemanth Yeddulapalli$^\$$, Prasad Calyam$^\$$, Saptarshi Debroy$^\ast$\\
$^\ast$City University of New York, $^\$$University of Missouri-Columbia\\ Emails: \textit{mmounesan@gradcenter.cuny.edu, lemusm@umsystem.edu, hygw7@missouri.edu, calyamp@missouri.edu,\\ saptarshi.debroy@hunter.cuny.edu}}

\maketitle
\begin{abstract}

In recent times, Volunteer Edge-Cloud (VEC) has gained traction as a cost-effective, community computing paradigm to support data-intensive scientific workflows. However, due to the highly distributed and heterogeneous nature of VEC resources, centralized workflow task scheduling remains a challenge. In this paper, we propose a Reinforcement Learning (RL)-driven data-intensive scientific workflow scheduling approach that takes into consideration: i) workflow requirements, ii) VEC resources’ preference on workflows, and iii) diverse VEC resource policies, to ensure robust resource allocation. We formulate the long-term average performance optimization problem as a Markov Decision Process, which is solved using an event-based Asynchronous Advantage Actor-Critic based RL approach. Our extensive simulations and testbed implementations demonstrate our approach’s benefits over popular baseline strategies in terms of workflow requirement satisfaction, VEC preference satisfaction, and available VEC resource utilization.
\end{abstract}


\section{Introduction}

Data-intensive scientific workflows in areas characterized by considerable on-demand 
resource needs 
and stringent security requirements (e.g., bioinformatics, high-energy physics, and healthcare), 
have 
traditionally been hosted by
cloud environments, 
thanks to the availability of resources,
advanced security protocols, and performance assurances through Service Level Agreements (SLAs)~\cite{mell2011nist} offered by such environments. 

\begin{figure}[t]
    \centering
    \includegraphics[width = \linewidth/2]{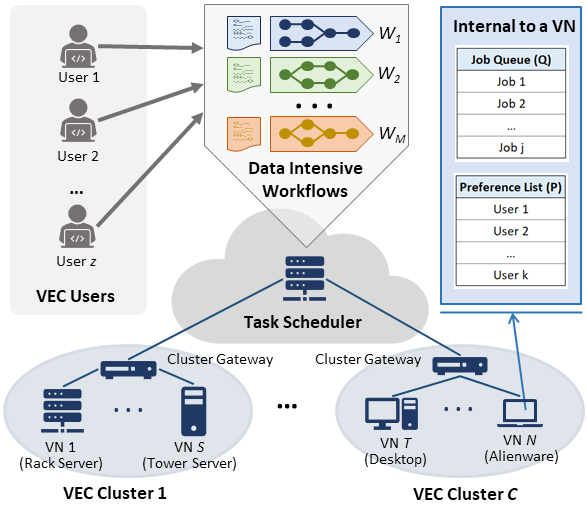}
    \caption{\footnotesize{Data-intensive workflow scheduling within a VEC environment }}
    \label{fig:vec-overview}
    \vspace{-6mm}
\end{figure}

However, processing such data- and resource-intensive workloads at cloud scale incurs substantial costs.
To address this, in recent times, ``volunteer edge-cloud" (VEC) computing has emerged as an alternative~\cite{seti,boinc}, harnessing distributed computing to provide cost-effective resources~\cite{pires2021distributed} for on-demand processing.
Figure~\ref{fig:vec-overview} illustrates an exemplary VEC environment that leverages the collective computational resources of VEC nodes (i.e., VNs) to process data-intensive workflows; thereby shifting the processing from centralized cloud infrastructures to the edge, where resources are more affordable and abundant, albeit diverse and geographically distributed.

These VNs can range from small devices (e.g., IoTs) to large systems (e.g., servers) that are owned and operated by individuals, laboratories, or organizations who willingly contribute 
them for collaborative computing. 

A central scheduler is designated to assign workflow tasks to available VNs that can satisfy workflows' quality of service (QoS) and security requirements without violating the diverse VEC resource policies.

While traditional cloud environments provide theoretically unlimited resources 
to fulfill workflow requirements 
within specific SLA bounds, VNs within a VEC environment, due to their heterogeneity in terms of resource capacity, intermittent availability, and diverse usage policies, may not always guarantee strict requirement satisfaction. 
Additionally, VNs belonging to specific research labs/facilities within institutions/universities form isolated VEC clusters, while being part of the same VEC environment. 
These clusters may prefer to host specific workflows or users (generating such workflows) in their VNs due to a variety of preferential reasons, such as workflow data type, reputation of the workflow users, and history of prior collaborations between the data and resource sites. 
Thus, unlike in cloud environments, task scheduling in VEC environments needs to not only satisfy workflow demands, but also accommodate VNs' preferences. This is on top of optimizing task execution and efficiently managing resource scalability like any other task scheduling strategy.
Most related literature within VEC ecosystem focuses on establishing trust between the resource providers and users~\cite{vectrust,alarcon2022vecflex}, while mostly using generic task scheduling. 
Therefore, 
management of workflow tasks, resource assignment, and ensuring workflow requirement satisfaction, while honoring VNs' preferences for users/workflows remain some of the central challenges for VEC resource management~\cite{mengistu19}.

Unlike traditional cloud and edge systems~\cite{papagianni2013optimal, shrimali2020multi, tang2019dynamic, effectDNN}, where resource allocation is typically formulated as an optimization problem and solved using sub-optimal heuristics, resources in VEC systems are complicated to manage, due to their highly decentralized nature and heterogeneity.   
A VEC environment, comprising of multitude of VEC clusters, 
suffer from: a) diverse resource usage policies that might not be well laid-out, b) unpredictable usage pattern leading to ever-fluctuating job queue, and c) untrusted configurations that are difficult to predict.
Thus, in many cases,
task schedulers are unable to ascertain complete information about the capabilities and status of VNs (e.g., availability, trustworthiness, security posture, job queue length) 
belonging to such diverse clusters.
Classical optimization based approaches, thus, are ineffective in the presence of such imperfect information and system variability. 

In recent times, Reinforcement learning (RL) based approaches are being proposed 
for decision-making
under uncertain and dynamic environmental conditions~\cite{shyalika2020rl}, while addressing security concerns~\cite{he2022trsutrl} amidst extreme environmental fluctuations. In general, RL can learn from interactions within the environment, even when faced with incomplete information about resource availabilities and task requirements. Over time, RL can adapt resource allocation policies based on the feedback received through rewards and penalties, effectively learning how to allocate resources in a way that maximizes system efficiency and task performance.

In this paper, we 
introduce an RL-driven task scheduling approach for assigning data-intensive workflow tasks to diverse VNs within a VEC environment.
Our approach takes into consideration workflow QoS specifications (i.e., {\em QSpecs}) and security specifications (i.e., {\em SSpecs}) to satisfy workflow requirements. At the same time, the proposed approach considers the long-term trustworthiness (i.e., {\em trust}), resource specifications (i.e., {\em RSpecs}), and user/workflow preferences of the VNs belonging to different VEC clusters in order to ensure robust resource allocation and with the aim to satisfy both workflow-centric and resource-centric needs.
Our approach uses the above mentioned factors to formulate a long-term average performance optimization problem.
The proposed approach piggybacks on our earlier research on workflow specification formalization~\cite{multi-cloud} and trust computation within VEC environment~\cite{alarcon2022vecflex,vectrust} to formulate the long-term average performance optimization problem.
To find the optimal solution, we reframe the problem as a Markov Decision Process (MDP), which is then solved using an event-based Asynchronous Advantage Actor-Critic (A3C) based RL approach. The solution is implemented as the centralized task scheduling strategy for assigning an incoming workflow task (considered an event) to an available and suitable VN within the VEC environment.

We validate the effectiveness of our proposed RL-driven approach through a comprehensive simulation and a VEC testbed implementation. 
For the evaluation, we implement real bioinformatics data analytic workflows from the SoyKB science gateway~\cite{soykb} which are typically executed at community cloud sites, and thus serve as ideal candidates for VEC adoption. 
Specifically, we implement two workflows, viz., PGen and RNA-Seq that have varied {\em QSpecs} and {\em SSpecs}~\cite{pgen}. The former is comparatively complicated workflow that performs extensive next-generation data sequencing analysis, while the latter is relatively simpler, designed for gene expression quantization using transcriptomics data. 

In order to add workflow diversity in terms of requirements, we also incorporate two synthetic workflows into the simulation, augmenting them with artificially generated {\em QSpecs} and {\em SSpecs}
that mimic typical bioinformatics workflows.
The simulation results demonstrate our RL-driven approach's success in delivering high worklfow requirement satisfaction and resource preference satisfaction for varying task arrival rates and number of available VNs within the environment. As an added benefit, the results also show that our RL-driven long term optimization strategy can ensure that more than $50\%$ of VNs' job queues are at least $50\%$ full at all times, for realistic values of task arrival rates; thus demonstrating our approach's efficiency in utilizing available VNs. Finally, we demonstrate that our RL-driven approach performs significantly better than other popular baseline volunteer resource scheduling strategies~\cite{seti,alarcon2022vecflex} in terms of requirement satisfaction, task rejection rate, and available VN utilization. We additionally implement our RL-driven scheduling solution on a VEC
environment testbed, built on the Nautilus Kubernetes cloud
platform~\cite{nautilus} and running real bioinformatics workflows. The implementation results confirm the claims from simulation results, thus demonstrating great benefits of our proposed solution.  

The remainder of this paper is organized as follow: Section~\ref{sec:background} presents the research background and related work. Section~\ref{Sec:sys_model} describes the system model and formulates the 
problem. 
Section~\ref{Sec:A3C} describes the proposed RL-based approach. Section~\ref{sec:evaluation} discusses evaluation.  
Section~\ref{Sec:conclusions} concludes the paper.

\vspace{-0.1in}
\section{Background and Related Work}
\label{sec:background}

In this section, we present an overview of VEC environments, challenges in VEC resource management, and the current state-of-the-art and knowledge gaps. 

\subsection{VEC computing ecosystem}

Figure \ref{fig:vec-overview} portrays the foundational framework of a typical VEC system, comprising of: 
VEC users submitting workflows with specific requirements, a centralized scheduler tasked with assigning the workflow tasks to VNs, and VNs belonging to VEC clusters with their local job queue and user/workflow preferences.    

The users, i.e., scientists and researchers, strive to efficiently and affordably execute data-intensive workflows through on-demand computational resources delivered via a VEC service, often handled by a cloud-native, centralized task scheduler. The scheduler 

orchestrates intricate logic to align submitted workflow requirements with the best-fit resources from the available VNs.
On the other hand, the VNs or the clusters the VNs belong to, suggest user/workflow preferences that the scheduler tries to accommodate when assigning workflow tasks.

The VNs encompass a diverse range of hardware, spanning from rack servers to desktops, and from laptops to GPU accelerators with varied computational capabilities. The specific hardware configuration of VNs is contingent upon the contributions made by individual researcher labs/institutions that act as volunteers, donating their equipment when not in use. Consequently, the VEC ecosystem embraces a heterogeneous collection of resources, accommodating the availability and capabilities of participating volunteers' hardware. This flexible and decentralized nature of VNs enable the ecosystem to leverage a wide array of computational 
resources, fostering a collaborative and distributed environment for data-intensive scientific workflow execution.

\subsection{Resources management in VEC environments}

Various mechanisms are proposed to address the scheduling challenges of 
heterogeneous VEC environments. Maheshwari et al. \cite{maheshwari18} propose a hybrid edge cloud model that supports latency-sensitive applications in urban areas, optimizing resource provisioning as per requirements. Galletta et al. \cite{galletta18} introduce the CESIO architecture, enhancing video content delivery quality within the same edge. Funai et al. \cite{funai14} suggest an ad-hoc model where devices with internet access act as local task distribution points (TDPs), inviting other users to participate. Mengistu et al. \cite{mengistu18, mengistu19} leverage idle home IoT devices to expand the volunteer resource pool. Inspired by these concepts, Ali et al. \cite{ali21} propose a fog-cloud based task distribution layer, bringing cloud services closer to end users through fog nodes. Sebastio et al. \cite{sebastio18} present a holistic volunteer cloud model that employs Ant Colony Optimization (ACO) to optimize task-resource assignments. Pandey et al.~\cite{vectrust} propose a trust-based mechanism for allocating computational resources. Alarcon et al.~\cite{alarcon2022vecflex} and Rodrigues et al. \cite{rodrigues18} use Particle Swarm Optimization (PSO) to dynamically assign users to volunteer resources. 

{\em Unlike these existing works, we take a holistic approach 
that performs long-term joint optimization of workflow requirements and VN preferences, while considering VN resource policies and long-term trust, using a black-box approach which is more practical, yet challenging to solve.}

\subsection{RL for distributed resource management}

Reinforcement Learning (RL), particularly the Actor-Critic method,
shows great promise in enhancing resource allocation, task scheduling, and overall system performance, especially in 
dynamic and black-box environments like VEC computing. 

Fu et al.~\cite{fu2020actor} propose an innovative Actor-Critic mechanism to manage offloading decisions and resource allocation in Mobile Edge Computing (MEC) environments. 
Similarly, Wei et al.~\cite{wei2017user} focus on optimizing user scheduling and resource allocation in heterogeneous mobile networks using a policy-gradient-based Actor-Critic approach.  
Shah et al.~\cite{shah2021joint} address network utility maximization in massive IoT environments by proposing a hierarchical deep Actor-Critic model for network management and resource allocation. 

Additionally, Chen et al.~\cite{chen2019learning} introduce an Actor-Critic method-based framework to optimize resource allocation in cloud data centers, targeting improved job execution latency and resource utilization. 
Meanwhile, Tathe et al.~\cite{tathe2018dynamic} focus on down-link Transmission for Long Term Evaluation Advanced (LTE-A) radio resource allocation, proposing an Actor-Critic based architecture to maintain QoS and user fairness amidst dynamic scheduling challenges. 
{\em These collective results demonstrate the effectiveness of the Actor-Critic approach in handling the dynamic and black-box nature of environments. Motivated by these outcomes, 
we pursue an Actor-Critic based approach, viz., A3C for task scheduling in VEC environments.
To the best of our knowledge, no such approaches exist that seeks to optimize resource allocation in volunteer computing environment, taking into consideration the requirements and preferences from both workflow and resource sides.}
 
\section{System Model and Problem Formulation}
\label{Sec:sys_model}
In this section, we describe the system model and formulate the optimization problem.

\subsection{System model}
\noindent The main components of our VEC system are as follows:

\vspace{0.05in}
\noindent\textbf{Workflows and tasks:} We define a set of \textit{m} workflows $\mathcal{W}=\{W_a, W_b, \ldots, W_m \}$ that uses the VEC environment. 
An instance of a workflow is referred to as a $task$. Each $task$ is a sextuple $(w, data, time, userID, QSpecs, SSpecs)$ 
representing the workflow, input data, submission time, user ID of the task submitter, QoS requirements, 
and security requirements of the specific task within the workflow, respectively. $QSpecs$ 
is a formalized and quantifiable way of representing a workflow task's
desired performance requirements 
that includes QoS metrics, such as throughput, latency, and response time.
Whereas, $SSpecs$ specifies a task's security requirements across certain (say $F$) security factors, as recommended by NIST SP 800E guidelines~\cite{force2013security, multi-cloud}. The 
level of each of these factors is set as High/Moderate/Low 
based on the NIST guidelines. 
Both $QSpecs$ and $SSpecs$ concepts are borrowed from the seminal work by Dickinson et al.~\cite{multi-cloud}, while the process of generating $QSpecs$ and $SSpecs$ for data-intensive workflow tasks can be found in~\cite{nguyen20-icdcn}.

\vspace{0.05in}

\noindent\textbf{VEC cluster:}
The VEC environment is composed of a collection of \textit{C} clusters denoted as \mbox{$\mathcal{C} = \{VEC_1, \ldots, VEC_{|\mathcal{C}|} \}$}, where each cluster consists of a varying number of VNs. From the perspective of the scheduler, VNs are considered as individual entities that operate independently. Thus, in the formulation and management of tasks, we consider VNs as distinct entities, each with its own characteristics and specifications. 

\noindent \textbf{VEC nodes (VNs):} 
We define a set of \textit{N} VNs $\mathcal{V}= \{VN_1, VN_2, ..., VN_N\}$ in the environment. Each $VN \in \mathcal{V}$ is another sextuple $(deviceID, RSpecs, P, config, T, Q)$ 

representing VN's identification number, resource specification, preference list, configuration, trust, and local queue, respectively. The resource specifications, denoted as $RSpecs$, define a set of factors that describe the security posture and usage policies of a VN, also adopted from~\cite{multi-cloud}. Additionally, the preference list $P$ is an ordered list 
of $\rho$ workflow users. As described earlier, the preferences 
can be based on a variety of factors, such as, workflow data type,
reputation of the workflow users, and history of prior collaborations between the data and resource sites.
The VNs exhibit heterogeneous configurations, yet VNs within the same cluster share common specifications in terms of guaranteed security measures and policies as well as a preference list. Furthermore, we consider a local job queue of maximum size $\Gamma_j$ for each $VN_j$.

\vspace{0.05in}

\noindent\textbf{Trust:} The trustworthiness of a $VN_j$  is denoted by a quantifiable trust metric $T_j$ and is defined as the level of consistency the VNs exhibit over time in terms of performance, agility, cost, and security (PACS) factors, as defined in~\cite{vectrust}. Given the voluntary nature of VEC resources, the VEC clusters may incidentally modify configurations, such as adjusting capacity or availability, or change security settings. Consequently, consistent provisioning of resources and configurations becomes indicative of reliable VNs.

\vspace{0.05in}

\noindent\textbf{Task assignment:}  
Depending on the the task requirements and VN availability, a task maybe accepted and assigned to a $\textrm{VN}$ or rejected. We use the symbol $\textrm{NULL}$ to represent rejection of a task. Let $g$ denote the assignment function that maps tasks into the elements of $\mathcal{V} \cup\{\textrm{NULL}\}$:

\begin{equation}
    g(t, task_i) = 
        \begin{cases}
            \textrm{VN}_j, &  \textrm{if } task_i \textrm{ is assigned to} \;\textrm{VN}_j\\[2ex]
            \textrm{NULL}, & \textrm{if } task_i\;\textrm{is rejected}\\
          
        \end{cases}       
\end{equation}
To quantify the quality of an assignment, we define a satisfaction score denoted by $\mathbb{S}$  that evaluates the assignment in terms of both tasks and VNs.

\subsection{Task satisfaction score}
The task satisfaction score measures the alignment of task's $QSpecs$ and $SSpecs$ with the resource configuration and $RSpecs$ of the assigned VN, respectively. It has two parts: 
\begin{itemize}[leftmargin=*]
    \item \textbf{QoS satisfaction score:} The QoS satisfaction score, i.e., $\textit{QSpecs}\mathbb{S}$ measures the alignment of task $QSpecs$ with the estimated performance that $VN_j$ offers. Let $WT$ and $Exe$ represent the estimated waiting time in the queue and estimated execution time of $VN_j$, respectively. Here, $WT$ in VN's queue ($Q$) is sum of the execution times of all the existing jobs in the queue. Thus,
    \begin{equation}
    WT(VN_j) = \sum_{i=0}^{|Q_j|} Exe(task_i, VN_j)
    \label{eq:localqueuetime}
    \end{equation}
where $|Q_j|$ stands for the size of $VN_j$'s queue. 
Then, we define the QoS satisfaction score $\textit{QSpecs}\mathbb{S}$ as:
\begin{equation}
    \textit{QSpecs}\mathbb{S}(task_i, VN_j) =
        \begin{cases}
            \frac{1}{1+e^{-\Delta_{i,j}}}, & \text{ if } \quad \Delta_{i,j} \geq 0 \\[2ex]
            \text{tanh}(\Delta_{i,j}), & \text{ if } \quad \Delta_{i,j} < 0 
        \end{cases}
\label{eq:qspec2}
\end{equation}
where $\Delta_{i,j}$ is the difference between the required latency of the task $QSpecs(task_i)$ and the estimated latency at $VN_j$: 
\begin{equation}
    \Delta_{i,j} = QSpecs(task_i) - WT(VN_j) -\textit{Exe}(task_i, VN_j)
\end{equation}    

\vspace{-0.05in}

\item \textbf{Security satisfaction score:} We define the security satisfaction score $SSpecs\mathbb{S}$ as the minimum distance between the required security level of the task $task_i$ and the offered security level by $VN_j$ across $F$ security factors, as described earlier. 
We propose to utilize a hard-security enforcement that does not allow an assignment of a task to a VN with a lower security level in any of the $F$ security factors. On the other hand, to manage resources more efficiently and avoid security over-provisioning, our proposed assignment strategy gives a lower security satisfaction score $SSpecs\mathbb{S}$ for assigning a task to a VN with strictly higher security level guarantees. For analysis, we assign the numerical values $1$, $2$, and $3$ to security categories Low, Moderate, and High, respectively. Thus,

\begin{mini}|s|
{f \in F}{ \delta^f_{i,j} }
{\label{eq:sspec}}
{SSpecs\mathbb{S} (task_i, VN_j) = } 
\end{mini}
where the distance function $\delta^f_{i,j}$ is defined over the $f$ security factor as follows:
    \begin{equation}
        \delta^f_{i,j} = 
        \begin{cases}
        \sqrt{\frac{3 - \left( RSpecs_{j}^{f}-SSpecs_{i}^{f} \right)}{3}}, & \text{ if} \quad RSpecs_{j}^{f}\geq SSpecs_{i}^{f}\\[2ex]
        0, & \text{ o/w}
        \end{cases}
    \end{equation}

 \end{itemize}
 
With $QSpecs\mathbb{S}$ and $SSpecs\mathbb{S}$ defined, we propose a joint QoS and security driven task satisfaction score $T\mathbb{S}$, where:
\begin{equation}
\scalemath{0.81}{
            T\mathbb{S}(task_i, VN_j) = 
            \begin{cases}
                0, &\text{if } \textit{SSpecs}\mathbb{S} (task_i, VN_j)=0\\[2ex]
                \begin{array}{@{}l}
                    c_1.SSpecs\mathbb{S}(task_i, VN_j)  \\
                    + c_2.QSpecs\mathbb{S}(task_i, VN_j),
                \end{array} &\text{ o/w}
            \end{cases} 
            \label{eq:satisfactionscore}
            }
\end{equation}  

\subsection{VN preference satisfaction score}
The VN preference satisfaction score, denoted by $VN\mathbb{S}$, is described as a function of an assigned users' rank in the VN's preference list. Specifically, we define $VN\mathbb{S}$ as a logarithmic function of $task_i$'s rank in $P_j$ (denoted by  $rank(task_i, P_j)$).

\begin{equation}
    VN\mathbb{S} =
    \begin{cases}
        1 - \frac{1}{6} \ln(rank(task_i, P_j)),  & \text{ if } task_i \in P_j\\[2ex]
        0,  & \text{ o/w} 
    \end{cases} 
\label{eq:vensatisfaction}
\end{equation}

\subsection{Overall satisfaction score}
The overall satisfaction score $\mathbb{S}$ of a task assignment to a VN is a function of $T\mathbb{S}$, $VN\mathbb{S}$, and trust $\mathbb{T}$ of the VN at the time of assignment. Thus: 

\begin{equation}
\scalemath{0.81}{
            \mathbb{S}(task_i, g(t, task_i)) = 
            \begin{cases}
                -b, &\text{ if } f(t, task_i) \text{ is }\textrm{NULL}\\[2ex]
                \begin{array}{@{}r}
                    a_1.\mathbb{T}_j(t).T\mathbb{S}(task_i, VN_j)  \\
                    + \ a_2.VN\mathbb{S}(task_i, VN_j),
                \end{array}  &\text{ o/w}
            \end{cases} 
            \label{eq:satisfactionscore}
            }
\end{equation} 
Here, constant $b$ incorporates dissatisfaction of rejecting a task.

\subsection{Formulating the optimization problem} 
We formulate the following optimization problem with the objective of jointly maximizing the average overall satisfaction score of the assignment strategy (over long-term), subject to the capacity of VNs' local queues and the hard-security requirement constraints explained earlier: 

\begin{maxi!}|s|[2]
 {}{ \lim_{T\rightarrow\infty} {\frac{1}{T} \sum_{t=1}^{T}} \sum_{i\in Tasks}\mathbb{S}(task_i, g(t, task_i)) }
 {\label{eq:optimization2}}{}
 \addConstraint{ |Q_j(t)| + \mathds{1}_{\{g(t, task_i) = VN_j\}} \leq \Gamma_j \quad \forall j \in [N] }{\label{queue}} 
 \addConstraint{SSpecs\mathbb{S}(task_i, g(t, task_i)) > 0 } {\label{security}} 
 \end{maxi!}
where \eqref{queue} ensures the assignment does not over-flow the local queues of the VNs, while \eqref{security} enforces a hard constraint on the security requirement of the submitted task which does not tolerate a lower than required security level of the assigned VN, and $\mathds{1}_{\{\cdot\}}$ denotes an indicator function. The optimization problem thus formulated is a multivariate NP-hard problem that demands the evaluation of all possible allocation permutations to determine the optimal solution. Due to the dynamic and black-box nature of VEC environment to the task scheduler, we choose a RL-driven approach to solve such complex optimization problem.

\begin{figure*}[t]
    \centering
    \includegraphics[width=\linewidth]{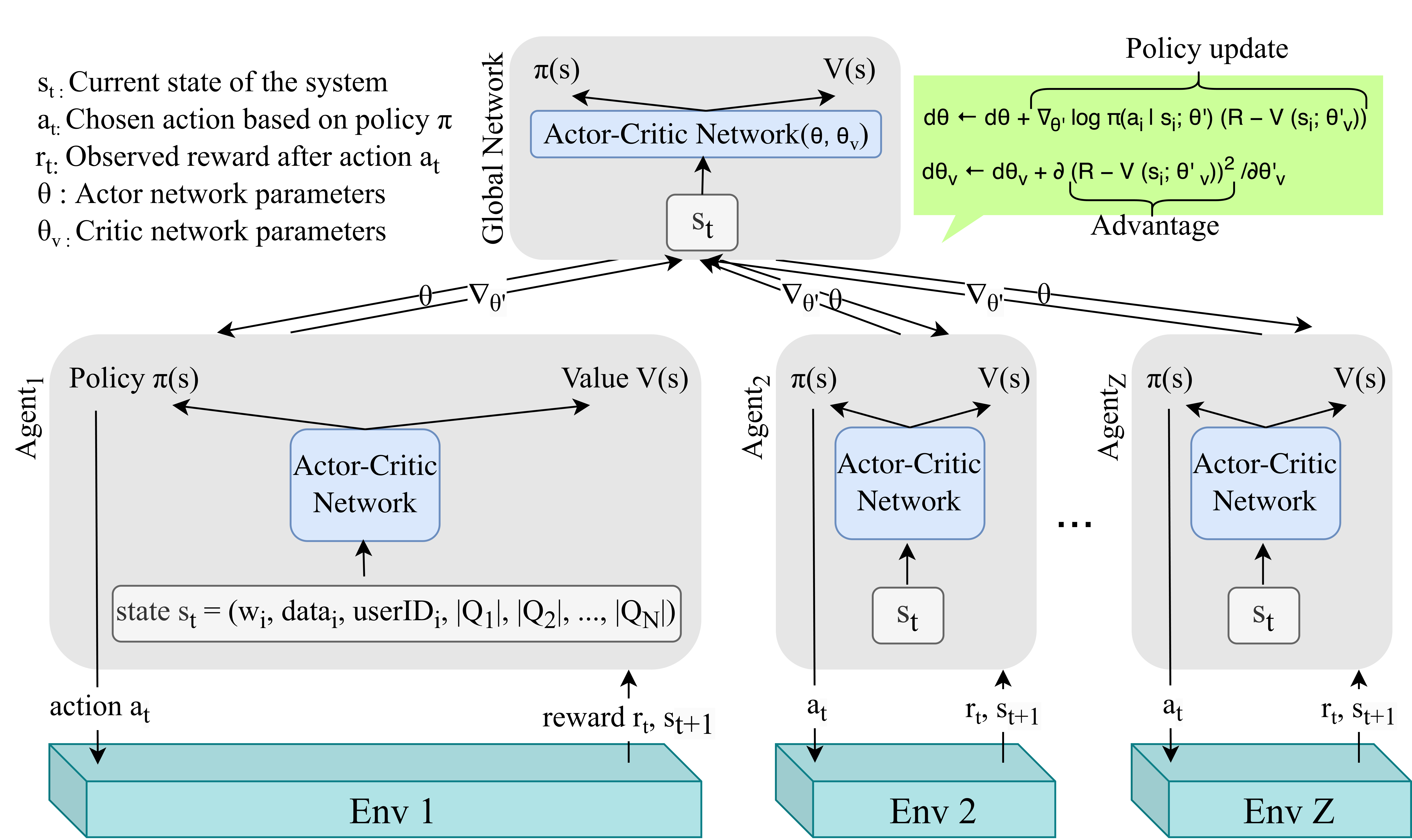}
    \caption{\footnotesize{The proposed A3C architecture 
    orchestrating multiple worker agents to concurrently interact with and learn from the environment
    }}
    \label{fig:a3c}
\end{figure*}

\section{Asynchronous Reinforcement Learning}
\label{Sec:A3C}
Here, we reframe the optimization problem 
as a Markov Decision Process (MDP) as it perfectly captures the dynamism of the VEC environment and introduces an event-based decision-making approach grounded in asynchronous deep reinforcement learning in order to solve the problem. More precisely,  we utilize Asynchronous Advantage Actor-Critic (A3C)~\cite{a3c} architecture to implement our scheduler strategy. This strategy is purposefully crafted to optimize the long-term average performance, as articulated in Eq.~\eqref{eq:optimization2}. We deploy parallel agents to learn an environment characterized by a finite set of states denoted as $\mathcal{S}$ and a finite set of actions denoted as $\mathcal{A}$. Next, we describe our A3C approach.

\subsection{Learning agents for the scheduler}
The combination of task information and task load within VNs' queues encapsulates a comprehensive representation of the system's state, fully discernible by our scheduler agent. This state encapsulates the specifics of the current task, as well as the status of local queues.

\noindent \textbf{States:} Let $\mathcal{S}$ denote the state space of the environment (i.e., our scheduler agent). The state of our scheduler agent at time $t$, denoted by $s(t) \in \mathcal{S}$, captures the particulars of the current submitted task $task_i$, including the task associated workflow ($w_i$), data ($data_i$), and userID ($userID_i$). 
Additionally, it consists of information about the VN's local queue status in relation to its respective load. For quantization, we consider four categories for a VN queue load based on the queue utilization: \textit{Low (L)}, \textit{Medium (M)}, \textit{High (H)}, and \textit{Full (F)}. We define queue utilization as the ratio of the number of tasks in the queue ($|Q_j|$) over the queue capacity ($\Gamma_j$) of the $VN_j$, and define the state of the $VN_j$ queue load at time $t$ by:  
\begin{equation}
    \mathcal{L}(Q_j(t)) = 
        \begin{cases}
            L, & \textrm{if } \quad \frac{|Q_j|}{\Gamma_j} \leq 0.3\\ 
            M, & \textrm{if } \quad 0.3 < \frac{|Q_j|}{\Gamma_j} \leq 0.6 \\
            H, & \textrm{if } \quad 0.6 < \frac{|Q_j|}{\Gamma_j} \leq 0.9 \\ 
            F, & \textrm{if } \quad \frac{|Q_j|}{\Gamma_j} > 0.9
        \end{cases}   
\label{eq:load}        
\end{equation}
Here, the specific values (i.e., 30\%, 60\%, and 90\%) represent different levels of queue utilization. However, the analysis holds true all other different quantization levels and values.  
Consequently, the observations space $\mathcal{S}$ of our scheduler agent is captured by:
\begin{equation}
\scalemath{0.84}{
        \begin{aligned}
            \mathcal{S} = \Big \{ s(t) = [\overbrace{w_i, data_i, userID_i}^{\textrm{Task}_i}, \overbrace{\mathcal{L}(Q_1(t)), \ldots, \mathcal{L}(Q_N(t))}^{\textrm{VNs  load}}] ~:~ \forall j \in [N],\\
            \mathcal{L}(Q_j(t)) \in \{L,M, H, F\}, \textrm{w}_i \in \mathcal{W} \Big \}
        \end{aligned}
    }        
    \label{eq:state}
\end{equation}

\noindent\textbf{Actions:}

The action space of the scheduler agent, denoted by $\mathcal{A}$, is a discrete action space.  At time $t$, the action $a(t)$ performed by the scheduler agent is to either reject the submitted task or to assign the task to a particular $VN \in \mathcal{V}$:
\begin{equation}
    \mathcal{A} = \{ a(t) ~:~ a(t) \in \mathcal{V} \cup \{\textrm{NULL}\} \}
    \label{eq:action}
\end{equation}

\noindent\textbf{System reward:}
The reward function $R(t)$ denotes the instant reward acquired following the transition from state $s(t)$ to state $s(t + 1)$ by executing the action $a(t)$. In our proposed A3C model, this reward function is realized as the satisfaction score in Eq.~\eqref{eq:satisfactionscore} of the allocation or the penalty of rejection:
\begin{equation}
    R(t) = \mathbb{S}(task_i, g(t, task_i)) 
   \label{eq:reward}
\end{equation}

\subsection{A3C network architecture and algorithm}
As shown in Fig.~\ref{fig:a3c}, our A3C~\cite{a3c} architecture comprises of two components: the actor network and the critic network. The actor network learns a policy $\pi$ that guides scheduler action selection, while the critic assesses the value of states, offering feedback for policy enhancement. A3C employs a parallelized approach by deploying multiple worker agents simultaneously, each operating within its own independent environment. This strategy fosters a diverse training experience and accelerates the learning process, particularly beneficial when handling larger observation spaces, such as ours as encountered when the number of VNs increases.

We design an offline learning algorithm (as shown in Algo.~\ref{alg:a3c}) for A3C driven task scheduling. 
In the initialization phase, the agents build actor and critic networks with random weights. Then the scheduler agent continuously interacts with the current environment and makes assignment decisions after each task submission. At the end of each episode, both actor and critic networks' weights are updated with a batch of experienced transitions. Our network is structured with a basic architecture, consisting of two fully connected layers, each with a feature size of 512 and 256, respectively. We've open-sourced the study's source code on GitHub~\cite{repo}.

\begin{algorithm}[htb]
\caption{A3C-based task scheduler training
}
\label{alg:a3c}
\parbox[t]{\dimexpr\linewidth-\algorithmicindent}{\textbf{Assume} global shared parameter vectors $\theta$ and $\theta_v$ for Actor and Critic networks, respectively}
\parbox[t]{\dimexpr\linewidth-\algorithmicindent}{
\textbf{Initialize} thread-specific parameter vectors $\theta'$ and $\theta_v'$ randomly}
\parbox[t]{\dimexpr\linewidth-\algorithmicindent}{\textbf{Output:} Global shared parameters $\theta$ and $\theta_v$ }
\begin{algorithmic}[1]

\For{episode=1 to $\mathcal{Z}$ (Total number of Episodes)}  
    \State \parbox[t]{\dimexpr\linewidth-\algorithmicindent-15pt}
    {Synchronize thread-specific parameters $\theta' \leftarrow \theta$ and $\theta_v' \leftarrow \theta_v$}

    \State Reset gradients: $d\theta \leftarrow 0$ and $d\theta_v \leftarrow 0$.
    \State Reset agent's state $s$
    \State Initialize empty episode buffer $s_1, a_1, r_1, \ldots, s_t, a_t, r_t$    

    \State $t \gets 0$
    \For{t=1 to T}
        \State // Assume $task$ is submitted at time $t$
        \State \parbox[t]{\dimexpr\linewidth-\algorithmicindent-15pt}{Update state $s_t$ based on the submitted $task$ and the status of VNs using Eq.~\eqref{eq:state}: $s(t) = [w, data, userID,\mathcal{L}(Q_1(t)), \ldots, \mathcal{L}(Q_N(t))]$}
        \State \parbox[t]{\dimexpr\linewidth-\algorithmicindent-15pt}{Input state $s_t$ to the Actor with weights $\theta'$ and generate action $a_t \in \mathcal{V} \cup \{\textrm{NULL}\}$ (Eq.~
        \eqref{eq:action}) according to policy $\pi(a_t|s_t; \theta')$}
        \State \parbox[t]{\dimexpr\linewidth-\algorithmicindent-15pt}{Receive reward $r_t = \mathbb{S}(task, g(t, task))$ (Eq.~\eqref{eq:reward}) and the next state $s_{t+1}$}
        \State Update the status of $task$ (allocated, rejected)
        \If{action $a_t$ not \textrm{NULL}}
            \State \parbox[t]{\dimexpr\linewidth-\algorithmicindent-15pt}{Add $task$ to the corresponding VN queue $Q_{a_t}$}
        \EndIf
        \State Append $(s_t, a_t, r_t)$ to the episode buffer $\beta$
    \EndFor

    \State \parbox[t]{\dimexpr\linewidth-\algorithmicindent}{Compute discounted rewards $\hat{R}_t$ for each time step $t$ and update $\theta'$ and $\theta'_v$}

    \State \parbox[t]{\dimexpr\linewidth-\algorithmicindent}{Perform asynchronous update of $\theta$ using $d\theta$ and of $\theta_v$ using $d\theta_v$.}
\EndFor
\end{algorithmic}
\end{algorithm}

\section{Evaluation} 
\label{sec:evaluation}
In this section, we evaluate the performance of our proposed task scheduling approach through an extensive simulation, followed by a testbed implementation on Nautilus Kubernetes cloud platform~\cite{nautilus}.

\subsection{Simulation environment} 
We begin by outlining 
the workflows used, their requirements, and the VEC environment.

\begin{itemize}[leftmargin=*]
\item \textit{Workflows:} In this work, we choose two high-throughput and typically cloud-native bioinformatics data analysis workflows in the SoyKB~\cite{soykb} science gateway developed for soybean and other related organisms. 

The complex PGen workflow is used to efficiently facilitate analysis of large-scale next generation sequencing (NGS) data for genomic variations. We also use a comparatively simpler RNA-Seq analysis workflow that is used to perform quantization of gene expression from transcriptomics data and statistical analysis to discover differential expressed gen/isoform between experimental groups/conditions. Given the frequency at which they are run (typically once or twice a week per user) and the total cost incurred for cloud adoption, they are ideal for VEC migration and an event-based task scheduling approach, such as ours. We also generate two synthetic  workflows in order to add diversity and scale to our workflow pool. 

Overall, the combined workflow tasks arrival rate to the task scheduler follows classic Poisson distribution.

\begin{table}[t]
    \centering
    \caption{Workflow $SSpecs$ for simulation}
    \begin{tabular}{c|ccccc}
        Workflow &  AC&CA&IA&SC&SI \\
        \hline
        PGen &H&H&H&L&L\\
        RNASeq & H&H&H&L&L\\
        Synthetic 1 & M&M&L&L&L\\
        Synthetic 2 & H&M&L&L&L\\
    \end{tabular}
    \label{tab:workflowsspec}
\end{table}

\begin{table}[t]
    \centering
    \caption{VNs' $RSpecs$ for simulation}
    \begin{tabular}{c|cccccc}
    RSpecs &Hardware &   AC&CA&IA&SC&SI\\
        \hline
        RSpecs 1&config1&H&H&H&M&L\\
        RSpecs 2&config1&H&H&H&L&L\\
        RSpecs 3&config1&H&H&H&M&M\\
        RSpecs 4&config1&H&M&L&L&L\\
        RSpecs 5&config2&H&H&H&M&M\\
        RSpecs 6&config2&H&H&H&L&L\\
    \end{tabular}
    \label{tab:venconfig}
\end{table}

\item \textit{SSpecs:} The details of the $SSpecs$ of PGen and RNASeq workflows are explained in \cite{nguyen20-icdcn}. $SSpecs$ for the synthetic workflows are simulated to add diversity to the $SSpecs$ pool. For this work, we only use $5$ out of $18$ security factors (as recommended by NIST) as they are the most relevant for VEC environments. These include: Access Control (AC), Security Assessment and Authorization (CA), Identification and Authorization (IA), System and Communication Protection (SC), and System and information Integrity (SI). 
The $SSpecs$ details are listed in Table~\ref{tab:workflowsspec}.

\item \textit{QSpecs:} Due to the scale-down of workflow datasize to fit the simulation scenario, simulated $QSpecs$ differ from real $QSpecs$ described in~\cite{nguyen20-icdcn}.
The determination of $QSpec$ for a workflow task involves assessing the average execution time when running that workflow with a specific data size on one of the standardized configuration. Additionally, we estimate the projected execution time of a task on a VN by analyzing data acquired from executing the same workflow with different data sizes within that specific configuration.

\begin{figure*}[!h]
    \centering
    \subfigure[]{\includegraphics[width=0.27\textwidth]{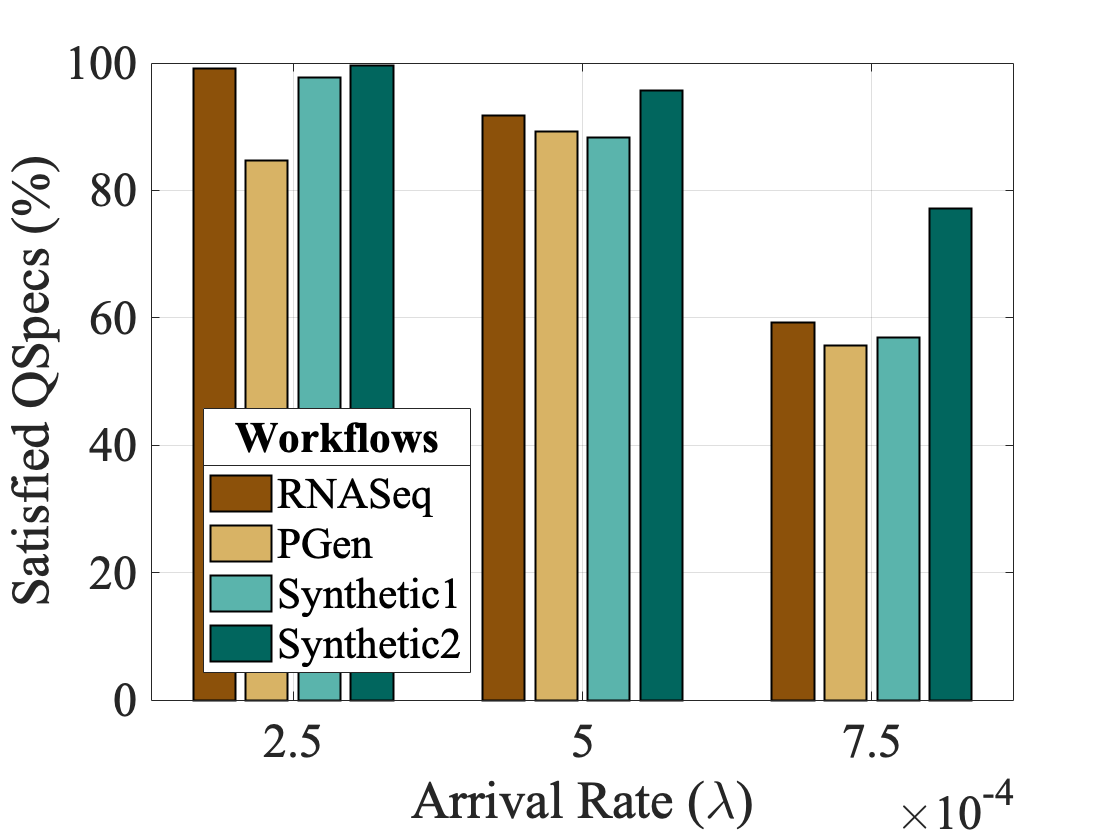}}\subfigure[]{\includegraphics[width=0.27\textwidth]{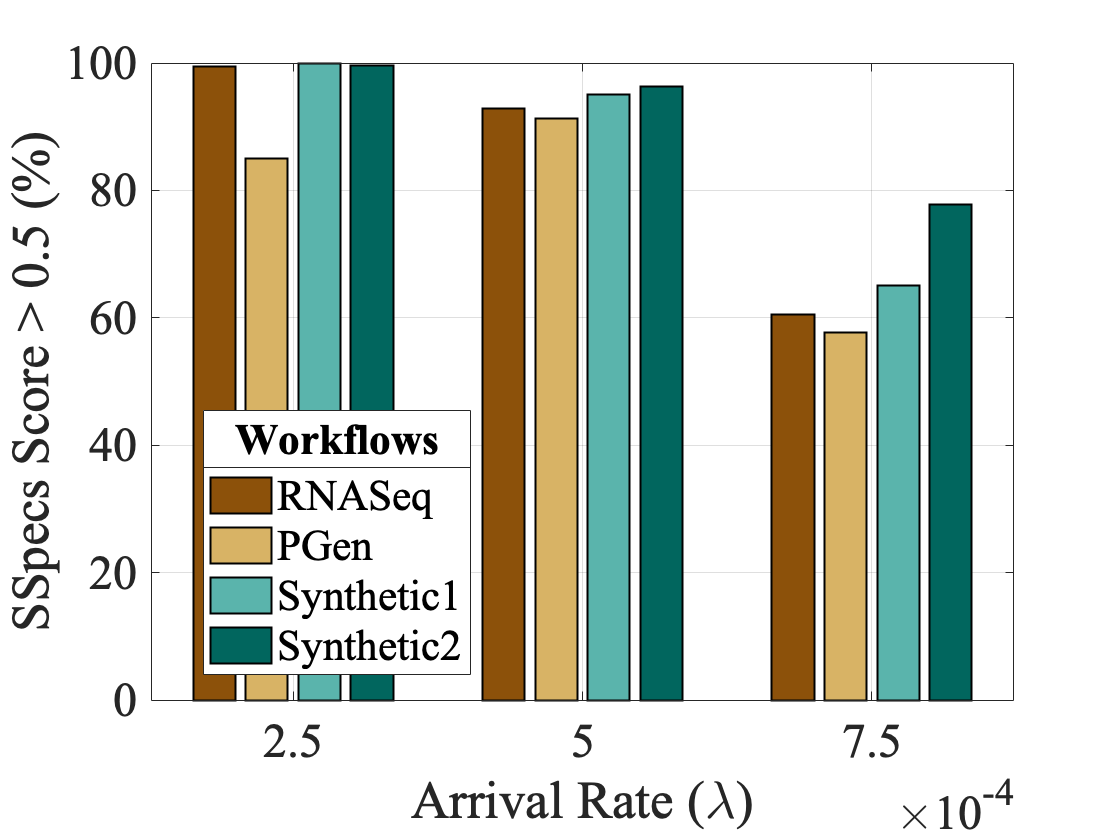}} 
    \subfigure[]{\includegraphics[width=0.27\textwidth]{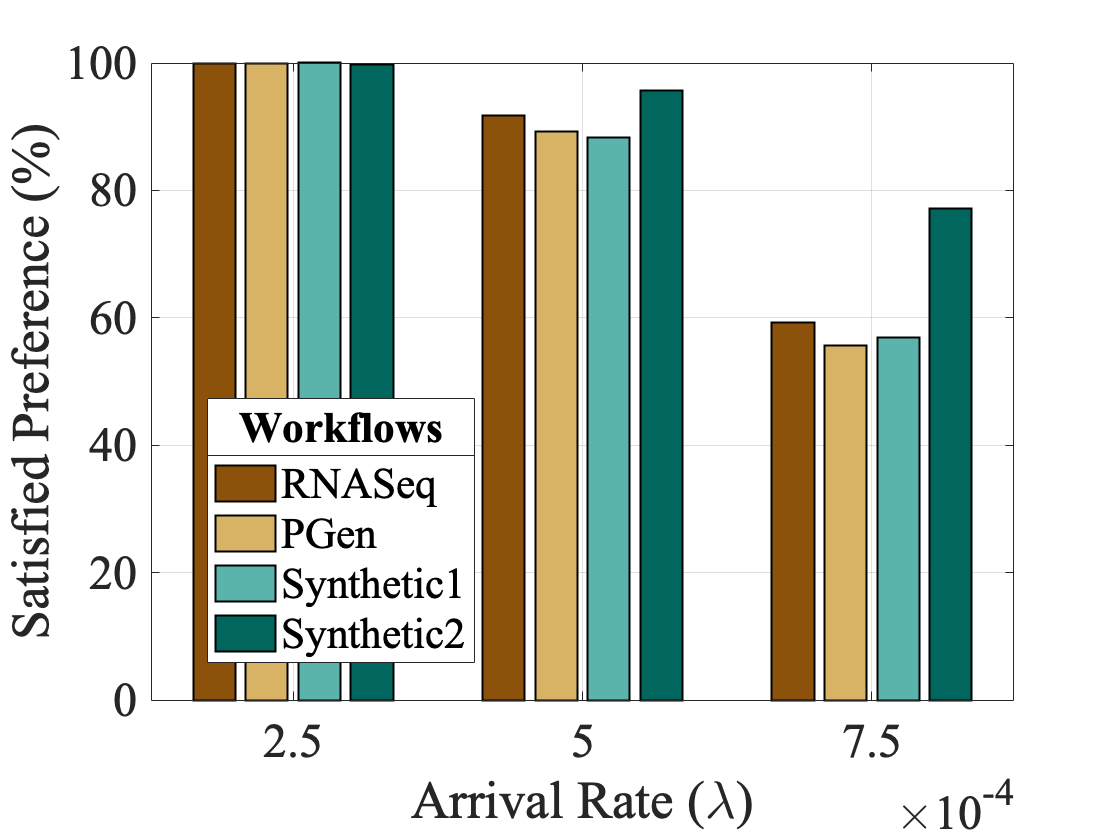}}
    \caption{\footnotesize{Requirement satisfaction for different task arrival rates}}
    \label{fig:foobar}
    \vspace{-0.1in}
\end{figure*}

\begin{figure*}[!h]
    \centering
    \subfigure[]{\includegraphics[width=0.27\textwidth]{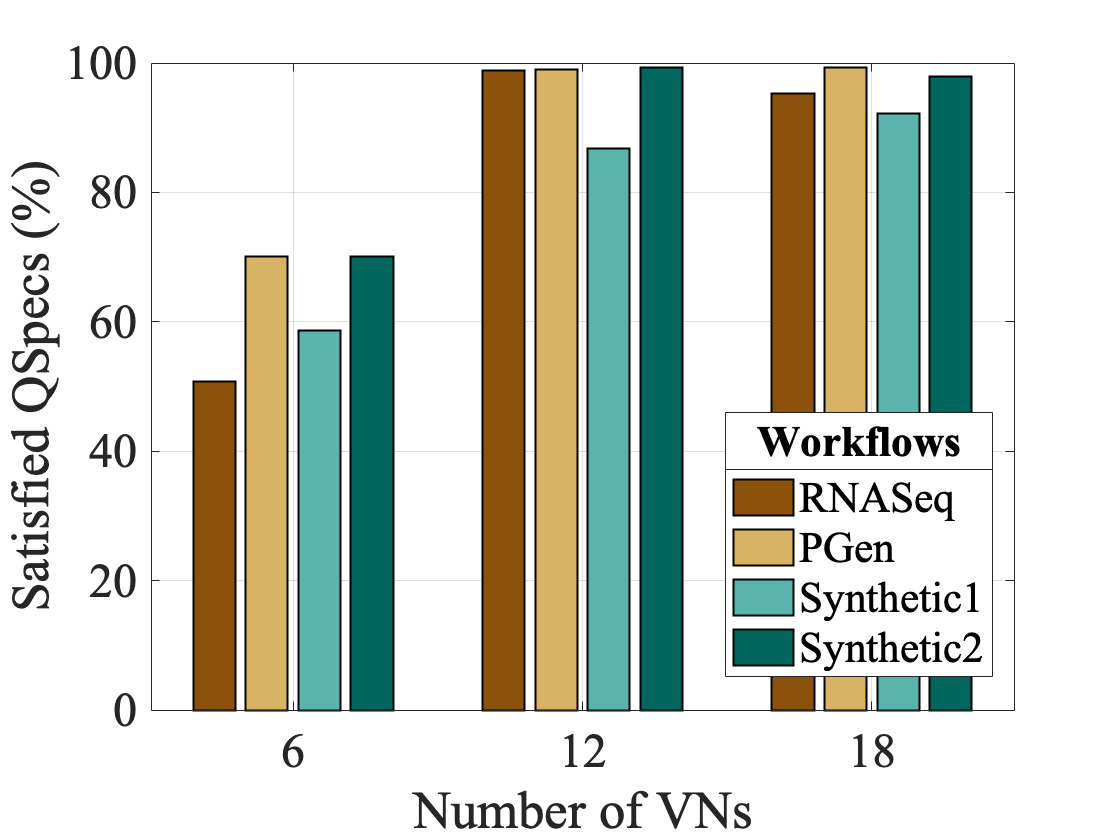}}\subfigure[]{\includegraphics[width=0.27\textwidth]{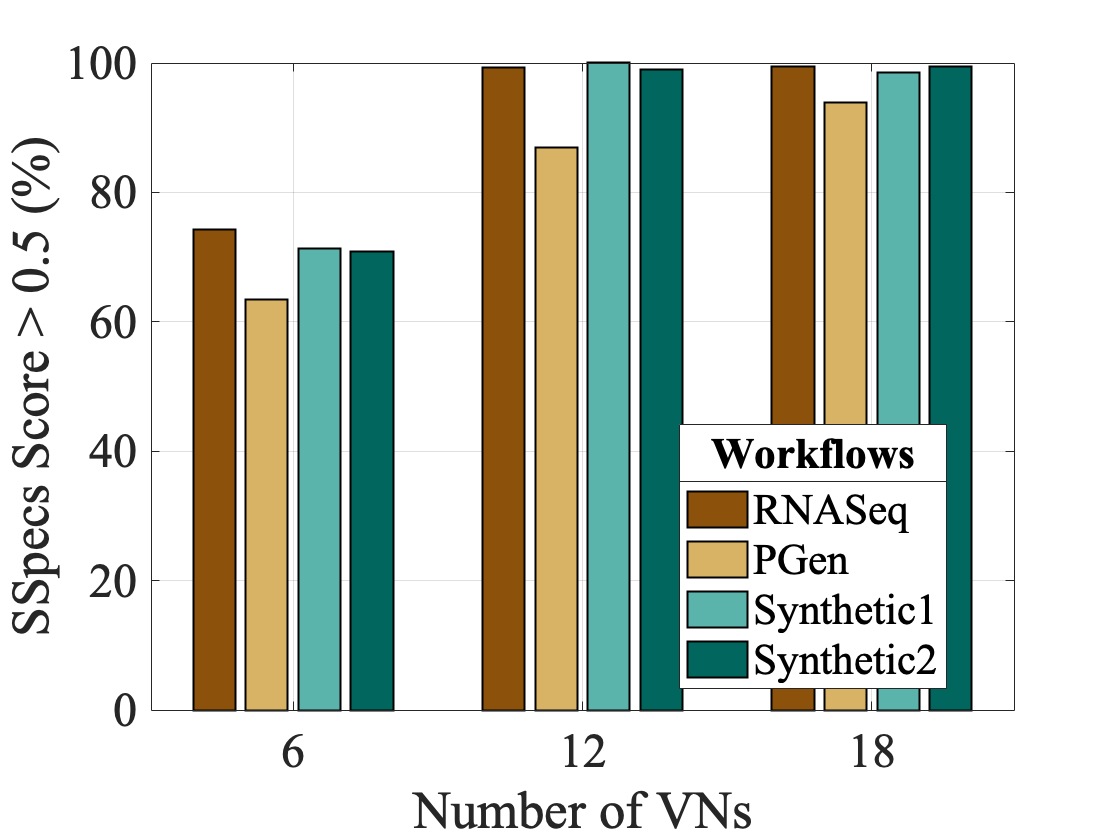}} 
    \subfigure[]{\includegraphics[width=0.27\textwidth]{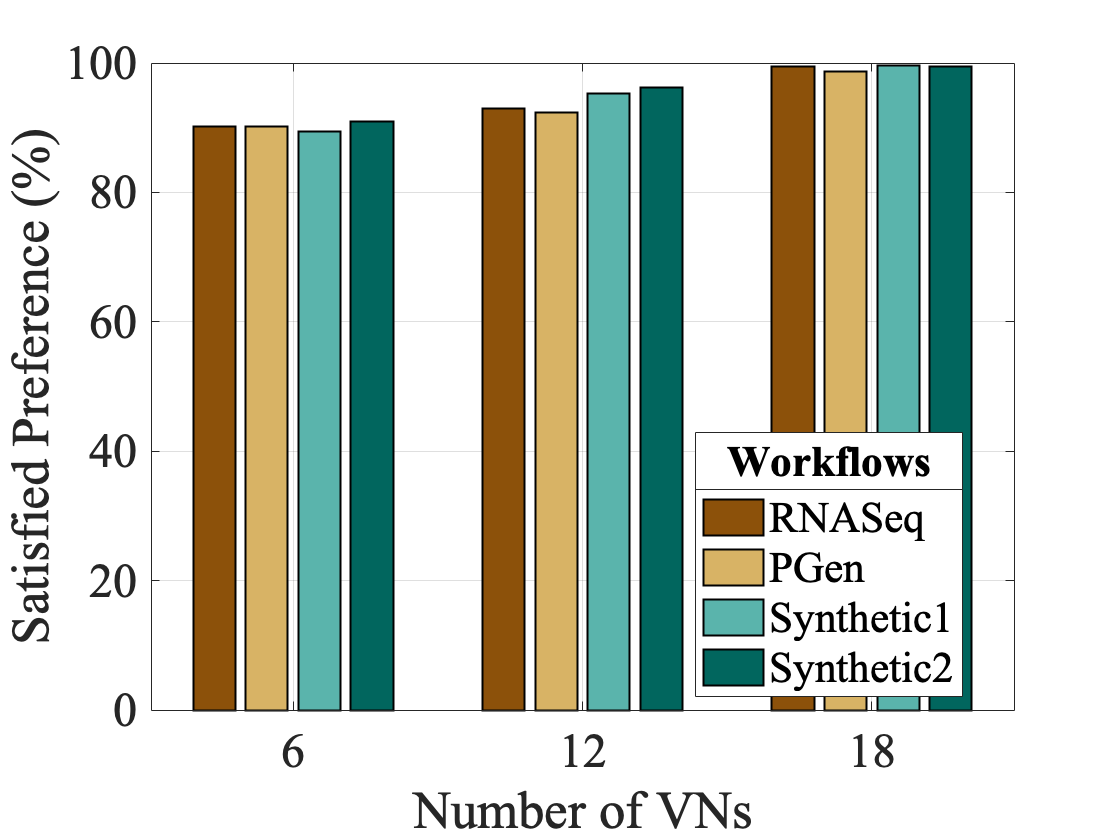}}
    \caption{\footnotesize{Requirement satisfaction for different numbers of VNs}
    \label{fig:foobar2}}
    \vspace{-0.2in}
\end{figure*}

\item \textit{VNs:}
With the objective of creating a diverse pool of VNs in terms of hardware and policy configurations, i.e., $RSpecs$, we simulate 6 $RSpecs$ configurations that are typical for a VEC environment comprising of lab based hardware as shown in Table~\ref{tab:venconfig}. Generating the configurations follows the security posture formalization and alignment technique described in~\cite{multi-cloud}.   
For the hardware, we use two distinct configurations that are typical for lab edge servers, they are: 1) PC with 32GB of RAM, Core i7 CPU with 2.8 GHz speed, and 2TB of disk space and 2) PC with 64GB of RAM, Core i9 CPU with 3 GHZ speed, and 4TB of disk space. 
The variations in $RSpecs$ and hardware configuration help us create a heterogeneous pool of 12 VNs (unless mentioned otherwise); 2 each for each combination described in Table~\ref{tab:venconfig}.
The workflow preference list of each VN is kept at $\rho = 5$ and is generated uniformly randomly. The maximum job queue capacity of each VN (i.e., $\Gamma$) is also kept at $5$, i.e., 
a new task assigned to a VN with $5$ jobs already in its queue is rejected.
Furthermore, for generating trust values of VNs, we use the principle of performance mismatch for trust estimation as described in VECTrust~\cite{vectrust}. 


\item\textit{Baseline approaches:} As baseline strategies for comparisons, we first simulate Particle Swarm Optimization (PSO) based scheduling  as deployed in `VECFlex' \cite{alarcon2022vecflex}.
Next, we use a completely `Random' scheme that assigns tasks to VNs in a randomized fashion without consideration on requirement satisfaction, with the goal of long term fairness. The next scheme is a Greedy-Random approach (`GR') that evaluates which VNs can satisfy the task QoS and security requirements and then randomly chooses one out of them. Finally, the Greedy-Best approach (`GB') always chooses the best VN in terms of requirement satisfaction. The greedy schemes are variations to state-of-the-art volunteer computing strategies such as~\cite{seti}.

\end{itemize}

\begin{figure*}[t]
    \centering
    \subfigure[VN with $RSpecs 1$]{\includegraphics[width=0.3\textwidth]{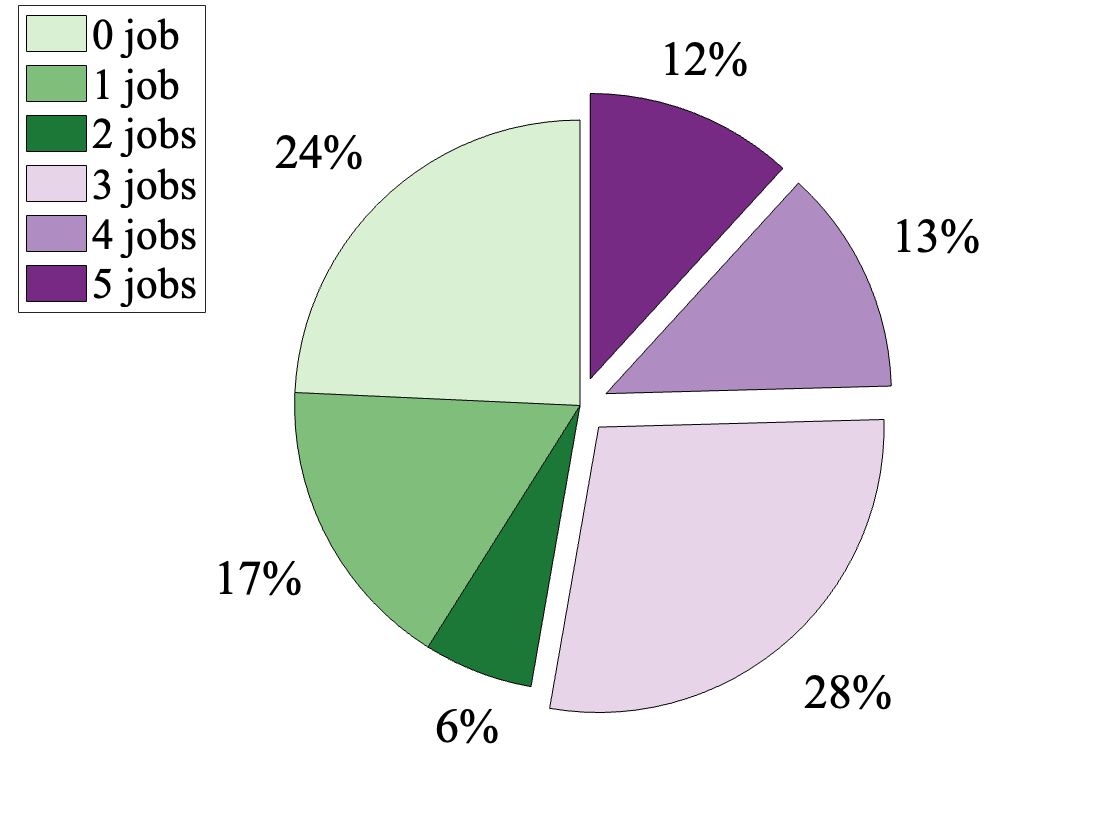}}\subfigure[VN with $RSpecs 6$]{\includegraphics[width=0.3\textwidth]{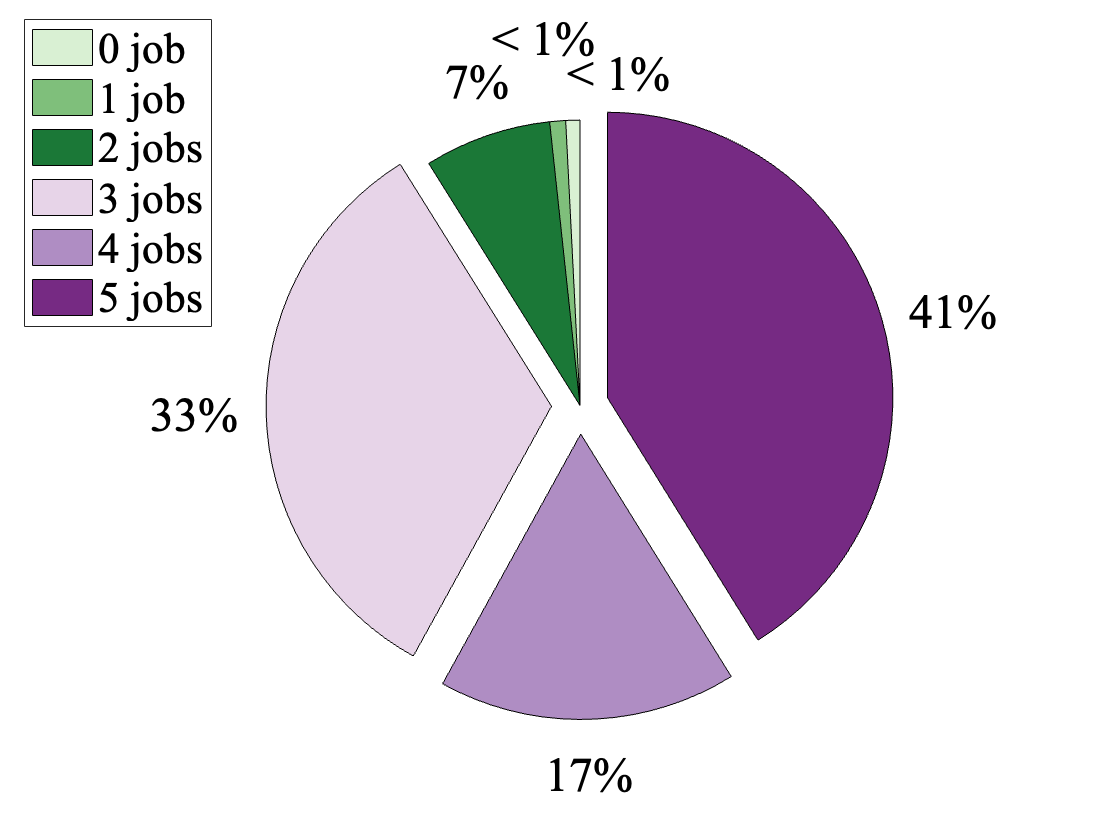}} 
    \subfigure[All VNs]{\includegraphics[width=0.3\textwidth]{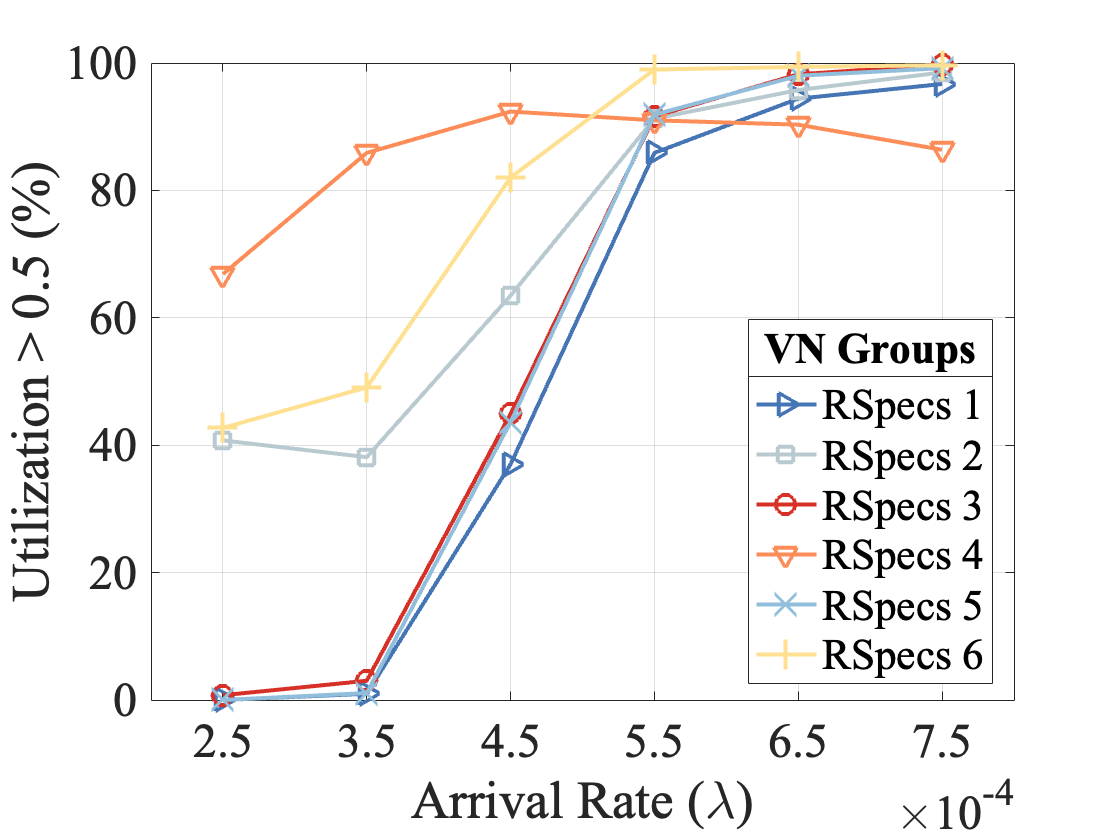}}
    \caption{Average utilization of VNs}
    \label{fig:util}
\end{figure*}

\begin{figure}[t]
    \centering
    \subfigure[]{\includegraphics[width=0.25\textwidth]{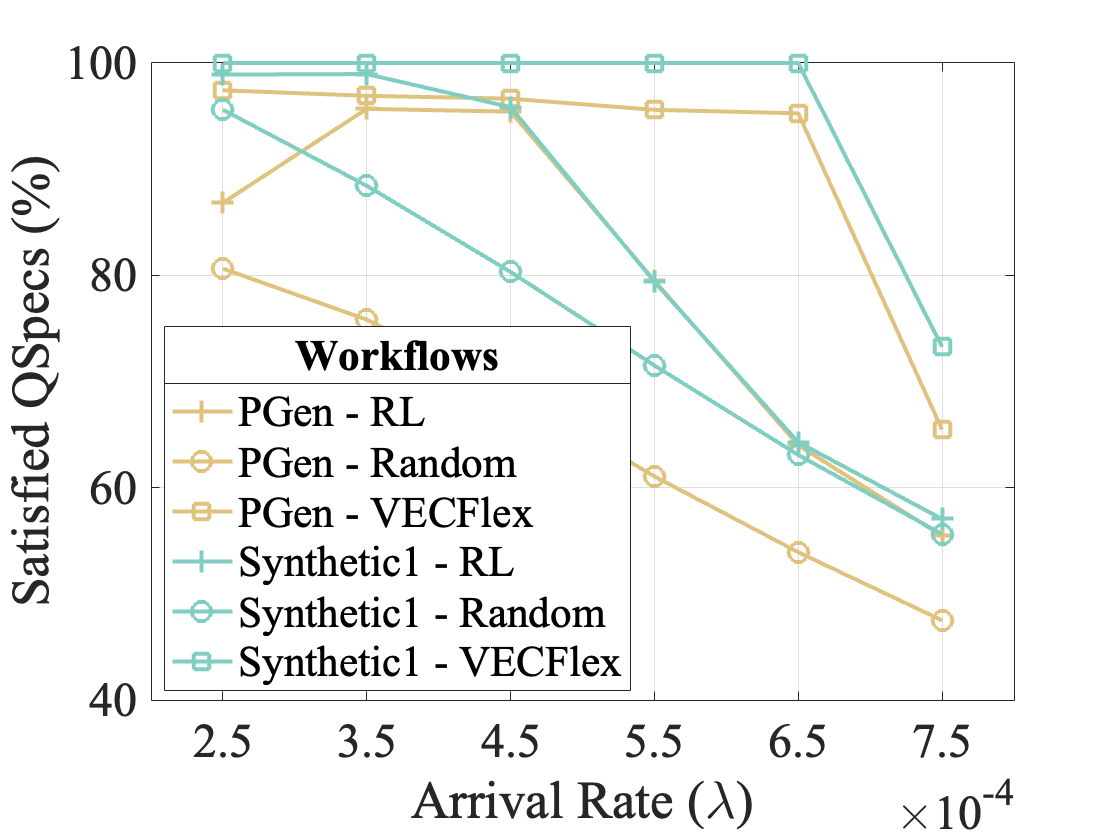}}
    \hspace{-0.2in}
    \subfigure[]{\includegraphics[width=0.25\textwidth]{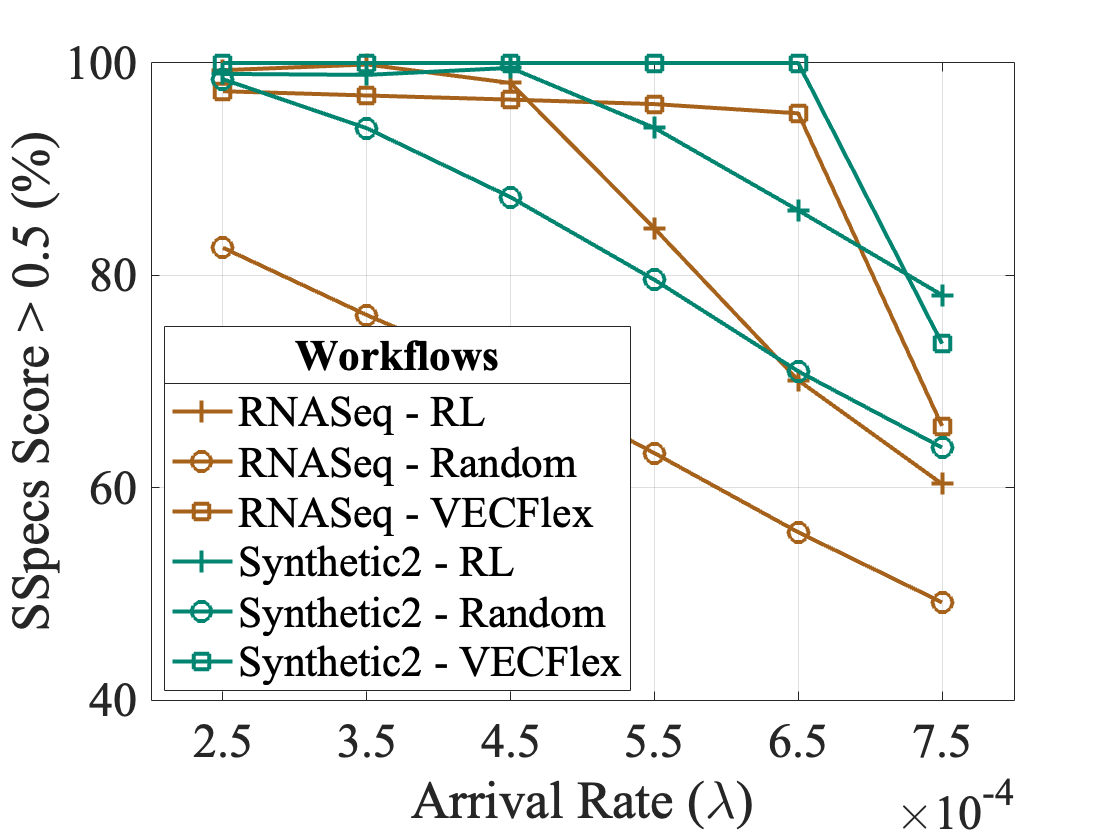}}
     \vspace{-0.1in}
    \caption{Workflow requirement satisfaction comparison}
     \label{fig:satis_comp}
     \vspace{-0.3in}
\end{figure}

\subsection{Simulation results}
Below, we discuss different aspects of the simulation results.

\noindent{\bf Requirement satisfaction:} In Fig.~\ref{fig:foobar}, we first show how our A3C based RL approach performs in terms of  satisfying task QoS and security requirements, and VN preference requirements, for different task arrival rates ($\lambda$). We observe that both $QSpecs$ and $SSpecs$ satisfaction performance is close to 100\% for lower job arrival rates. However, at very high $\lambda$, workflow satisfaction goes down due to high competition among workflows for limited VNs. Overall, we can observe that our proposed RL-driven task scheduling ensures requirement satisfaction
, for more than $50\%$ of the workflows. Fig.~\ref{fig:foobar2} shows requirement satisfaction performance against varying number of available VNs. We observe that both $QSpecs$ and $SSpecs$ satisfaction improve with more VNs in the environment as with more VNs, the probability of finding VNs with the right $RSpecs$ to match workflow requirements increases. It is interesting to observe that the synthetic workflows have higher probabilities of requirement satisfaction than PGen and RNASeq as the latter ones have stricter $SSpecs$ to satisfy. 

\noindent{\bf Average utilization:} In Fig.~\ref{fig:util}, we seek to ascertain the performance of our proposed approach in terms of average utilization of available VNs across the VEC environment. Figs.~\ref{fig:util}(a) and (b) show the percentage of different levels of utilization of two specific VNs (characterized by their $RSpecs$) for the entire duration of the simulation. Fig.~\ref{fig:util}(a) shows a VN with $RSpecs1$ which has the lowest average utilization over the simulation period. The figure shows that even for the most under-utilized VN, the job queue is more than $50\%$ full (i.e., with at least 2 jobs) for more than half the time. The utilization performance is even more impressive for the VN which is most utilized (i.e., VN with with $RSpecs6$), as shown in Fig.~\ref{fig:util}(b). 

The average performance of all VNs (with all $Rspecs$) for different job arrival rates (i.e., $\lambda$) is shown in Fig.~\ref{fig:util}(c) which shows that even with lower $\lambda$, many VNs are more than $50\%$ full for more than $50\%$ of the time. 

\noindent{\bf Satisfaction comparison:}
Next in Fig.~\ref{fig:satis_comp}, we compare our proposed approach (i.e., `RL') against two of the baseline approaches (i.e., `Random' and `VECFlex') in terms of task requirement satisfaction. Overall, the comparisons are carried out by running the simulation over $50$ times, each with different sets of workflow demands. 
In Fig.~\ref{fig:satis_comp}(a), we show the percentage of PGen and Synthetic1 workflows whose $QSpecs$ are satisfied by our RL-driven scheme versus Random and VECFlex. We observe that our our RL scheme follows VECFlex's greedy approach closely for smaller $\lambda$ values. Further, for all values of $\lambda$ and irrespective of the workflow, RL performs significantly better than Random assignment, even though the latter is designed for resource fairness. We see that $QSpecs$ satisfaction performance for Synthetic1 is better than PGen as the latter has stricter $QSpecs$ requirement. Fig.~\ref{fig:satis_comp}(b) also demonstrates our RL scheme's superiority over Random with RL delivering more than $0.5$ $SSpecs$ satisfaction score for almost $100\%$ of the workflows, especially for lower $\lambda$. Although the percentage of workflows with $SSpecs$ satisfaction score of at least $0.5$  decreases with larger $\lambda$, RL continues to perform better than Random. Comparing RL and VECFlex, it is evident that RL performs on par for smaller $\lambda$, but outperforms VECFlex as $\lambda$ approaches larger values.

\begin{figure}[t]
    \centering
    \subfigure[]{\includegraphics[width=0.25\textwidth]{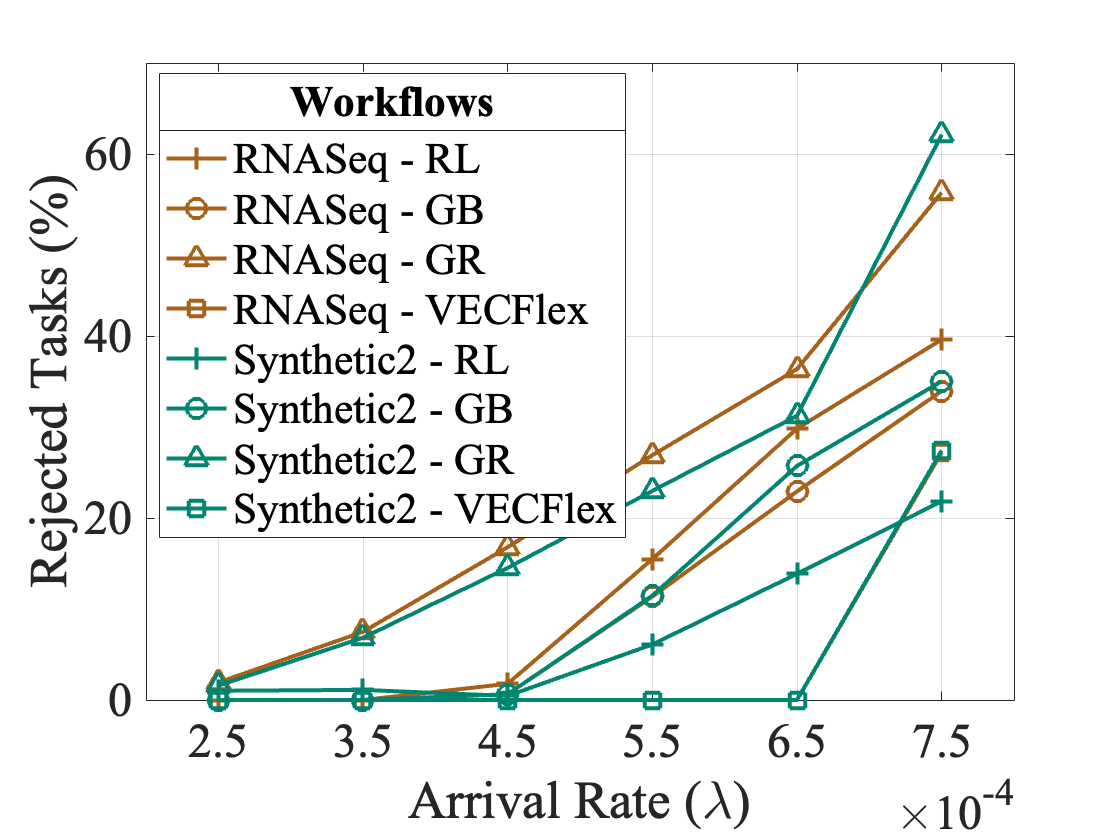}}
    \hspace{-0.2in}
    \subfigure[]{\includegraphics[width=0.25\textwidth]{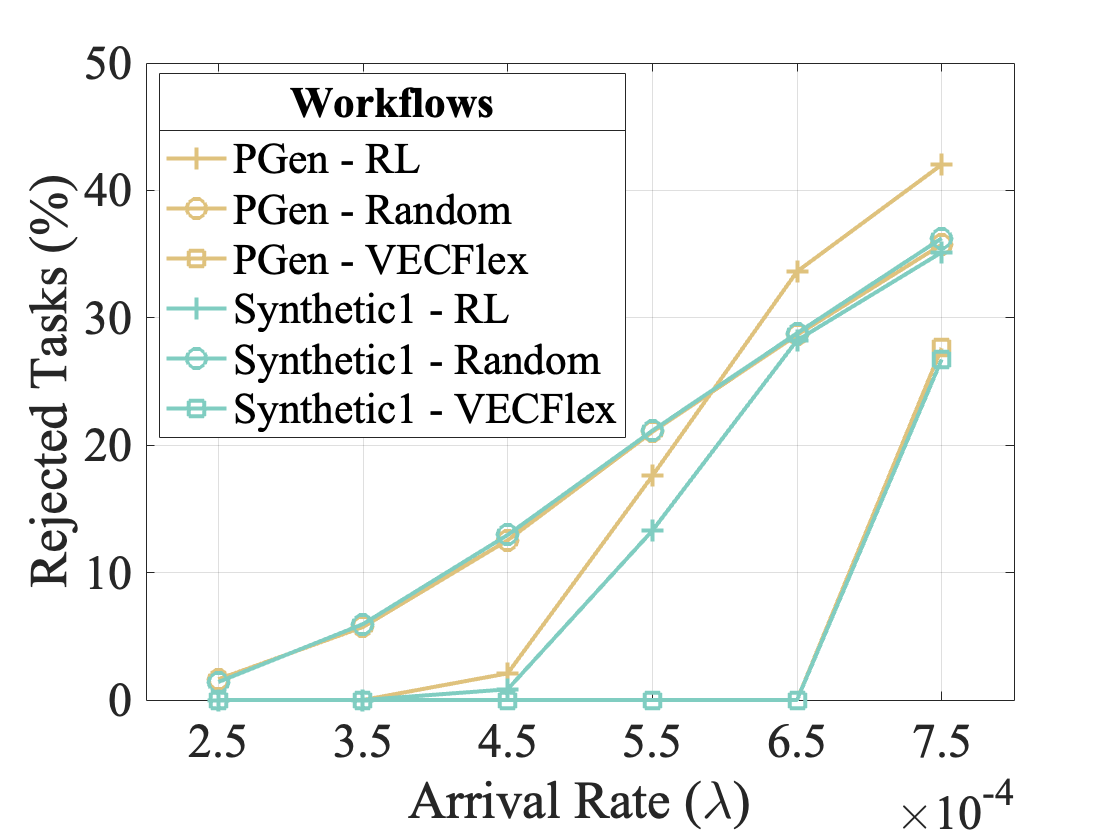}}
    \vspace{-0.1in}
    \caption{Rejection rate comparison}
    \label{fig:reject_comparison}
    \vspace{-0.2in}
\end{figure}

\noindent{\bf Job rejection rate comparison:}
In Fig.~\ref{fig:reject_comparison}, we compare the task rejection rates for our proposed RL-driven approach against other baseline strategies. Fig.~\ref{fig:reject_comparison}(a) demonstrates that for different $\lambda$, VECFlex and RL perform considerably better than GB and GR strategies for Synthetic2 workflow. However, for RNASeq, VECFlex and GB performs better than RL, while all perform better than GR. 

RL's better performance for Synthetic2 workflow can be explained by the computational complexity of this workflow; as RL model penalizes resource over-provisioning, it reserves resources for computationally intensive workflows, such as Synthetic2. The same reasoning justifies RL outperforming VECFlex for larger $\lambda$. However, for less intensive workflows such as RNASeq, GR and VECFlex perform better. 
In Fig.~\ref{fig:reject_comparison}(b), we compare the rejection rates of PGen and Synthetic1 workflows. 
As the PGen is computationally more demanding and has larger input data compared to Synthetic1, 
it results in higher rejection rates. On the other hand, Synthetic1 can be processed by almost all the VNs since its $SSpecs$ are less stringent. However, for both workflows, we can observe that RL performs comparable to VECFlex while significantly better than Random for most $\lambda$.

\begin{figure}[t]
\centerline{\includegraphics[width=0.4\linewidth]{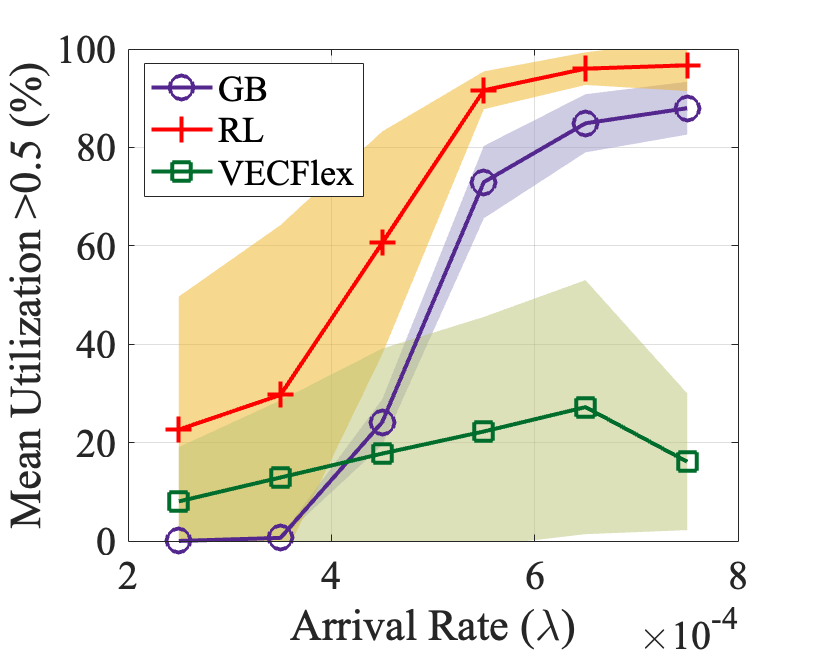}}
\caption{Mean VN utilization comparison}
\label{fig:rejectgreedy}
\end{figure}

\noindent{\bf Utilization comparison:}
Finally, in Fig.~\ref{fig:rejectgreedy}, we compare mean VN utilization of our RL-driven approach against GB and VECFlex. Here, we only consider those VNs whose queues are at least $50\%$ full (i.e., having more than 2 jobs in their queues) for the entire duration of the simulation. Here the lines represent the mean values with the shaded region representing the standard deviation for each data point of $\lambda$. We observe that for different values of $\lambda$, on average, RL performs better than GB and VECFlex, with RL performing significantly better than both for lower $\lambda$ values. This demonstrates our RL-driven approach's benefits in judiciously assigning workflows to VNs that best match workflow requirements with the VN policies. 

\subsection{Testbed implementation and results}
We implement our RL-driven scheduling solution on a VEC environment testbed, built on the Nautilus Kubernetes cloud platform~\cite{nautilus}, a specialized platform optimized for cloud-native applications and orchestrated containerized processes. The core components of the system consist of the proposed scheduler and the VEC environment.
The scheduler, featuring a robust backend, integrates a database system and a dedicated service for efficient task scheduling. It is hosted on containers created from the \textit{golang:1.20} Docker image~\cite{goland}. Communication occurs over ports 8080 and 3306 for backend and database services, respectively. The VNs, are constructed using the latest \textit{Go} Docker image and are equipped with a suite of bioinformatics tools and software, capable of running RNASeq workflow.

Each VN allocates specific resources, ranging from 4 CPUs and 8 GB of RAM to more powerful configurations, ensuring optimal performance for bioinformatics tasks. The entire networking infrastructure of the system is managed through Kubernetes services and ingress, guaranteeing secure and encrypted communication channels. 

For evaluation, we deploy RNASeq workflow with \textit{SSpecs} outlined in Table~\ref{tab:workflowsspec}, on 6 VNs with $RSpecs$ outlined on Table~\ref{tab:venconfig}. Due to the high cost of cloud services, we scaled down the size of workflows to maximum 1 GB, and evaluated the performance of the proposed RL-driven scheduling strategy over 1 hour for different task arrival rates (i.e., $\lambda$). It worth mentioning that the arrival rates are scaled in proportion to the new data sizes for the implementation.
Table~\ref{tab:testbed results} summarizes the results 
for key performance metrics, such as `Satisfied $QSpecs$', `$SSpecs > 0.5$', `Satisfied preference ($\%$)', `Rejected Tasks ($\%$)', and `Mean Utilization $\ge$ 0.5'. We observe that workflow and VN satisfactions values are close to $100\%$ with lower values of $\lambda$ and stays above $60\%$ even for larger arrival rates.
The testbed results thus corroborate the simulation findings demonstrating high effectiveness and efficiency of our proposed scheduling solution. 

\begin{table}[h]
\scriptsize
    \centering
    \caption{Scheduling performance for different $\lambda$}
    \begin{tabularx}{\columnwidth}{@{\extracolsep{\fill}} 
    c *{5}{c} }
        \hline
        $\lambda$
        & 
        \makecell[c]{Satisfied $QSpecs$(\%)} 
        & 
        \makecell[c]{$SSpecs$ $>0.5$(\%)\\}
        &
        \makecell[c]{Satisfied Preference(\%)}
        & 
        \makecell[c]{Rejected Tasks(\%)}
        &
        \makecell[c]{Mean Utilization $\geq$0.5(\%)}
        \\
        \hline
        0.02 & 94.82 & 98.27 & 98.27 & 1.72 & 2.58\\
        0.03 & 96.55 & 98.27 & 98.27 & 1.72 & 1.34\\
        0.04 & 86.20  & 86.20 & 89.65 & 10.34 & 2.90\\
        0.05 & 87.93  & 89.65 & 89.65 & 10.34 & 1.86\\
        0.06 & 64.01  & 66.56 & 64.01 & 33.43 & 5.70\\
        0.07 & 62.06  & 62.06 & 62.06 & 37.93 & 5.55\\
        \hline
    \end{tabularx}
    \label{tab:testbed results}
\end{table}


\section{Conclusions}
\label{Sec:conclusions}
In this paper, we introduced an A3C based RL-driven approach to data-intensive workflow task scheduling for VEC environments. We showed how our solution not only considered workflow QoS and security requirements, but also took into account diverse VEC resource policies dictated by various clusters (i.e., universities/labs/institutions) and their user/workflow preferences. Using extensive simulations and testbed implementation, we demonstrated how our proposed solution performed significantly better than other baseline strategies in terms of requirement satisfaction, task rejection rate, and available VN utilization.

\bibliography{references}
\bibliographystyle{ieeetr}

\end{document}